\documentclass[preprint]{ptephy_v1}
\usepackage{graphicx}
\usepackage{amsmath,amssymb}

\Year{2015}

\begin{document}

\title{
Angular momentum projected multi-cranked configuration mixing
for reliable calculation of high-spin rotational bands}

\author{Mitsuhiro Shimada, Shingo Tagami and Yoshifumi R. Shimizu}
\address{Department of Physics, Graduate School of Science,
Kyushu University, Fukuoka 812-8581, Japan}

%\date{\today}

\begin{abstract}

By employing the angular momentum projection technique we propose a method
to reliably calculate the quantum spectrum of nuclear collective rotation.
The method utilizes several cranked mean-field states with different
rotational frequencies and they are superposed in the sense of
the configuration mixing or the generator coordinate method,
after performing the projection;
the idea was originally suggested by Peierls-Thouless in 1962.
It is found that the spectrum as a result of the configuration mixing
does not essentially depend on chosen sets of cranking frequencies
if the number of mean-field states utilized in the mixing
is larger than a certain small value.  We apply this method to
three examples employing the Gogny D1S effective interaction
and show that it is useful to study high-spin rotational bands
by means of the angular momentum projection method.

\end{abstract}

%\pacs{21.10.Re, 21.60.Ev, 23.20.Lv}

\maketitle

%------------------------------------------------------------------------------
\section{Introduction}
\label{sec:intro}

The rotational motion in deformed nucleus is
a typical symmetry-restoring collective motion
and it appears as the rotational band, i.e., a group of
eigenstates generated from one deformed intrinsic state~\cite{BM75}.
In fact a nice rotational spectrum with $I^\pi=0^+,2^+,4^+, \cdots$,
can be obtained for the ground state band of even-even nuclei
by the method of angular momentum projection from
an axially-symmetric deformed state.
The angular momentum projection is a standard quantum mechanical technique
to restore the rotational invariance broken in the mean-field approximation,
and is suitable to microscopically describe
the nuclear collective rotation~\cite{RS80}.
However, this technique has not been very commonly utilized
especially for the description of the high-spin states except for
the works by the Projected Shell Model approach, see e.g. Ref.~\cite{HS95}.
In the present work we propose a method to reliably calculate the high-spin
rotational band based on the angular momentum projection method.

A standard method to describe the high-spin states
is the mean-field approximation combined with
the cranking model~\cite{RS80}; namely the Hamiltonian transformed into
the uniformly rotating frame (the so-called Routhian),
\begin{equation}
 H'= H-\omega_{\rm rot}J_y,
\label{eq:routh}
\end{equation}
is considered instead of the original Hamiltonian $H$.
Here the quantity $\omega_{\rm rot}$ is the rotational (or cranking)
frequency and $J_y$ is the angular momentum operator around the rotation-axis
(here chosen to be the $y$-axis).
It was first introduced by Inglis to calculate the moment of inertia
of the ground state rotational band~\cite{Ing54} treating the cranking
term $-\omega_{\rm rot}J_y$ as a lowest order perturbation, i.e.,
at the $\omega_{\rm rot}\to 0$ limit.
It was extended to the finite frequency to understand
various phenomena at high-spin states
as the selfconsistent cranking model~\cite{BMR73}
or the cranked shell model~\cite{BF79},
see e.g. Refs.~\cite{VDS83,GHH86,SW05}.
Recently its three-dimensional version,
i.e., the so-called tilted-axis-cranking~\cite{Fra93},
was also successfully applied, see e.g. Ref.~\cite{Fra01}.
The basic idea is that the selfconsistent mean-field state obtained
by minimizing the Routhian in Eq.~(\ref{eq:routh})
depends on the rotational frequency, $|\Phi_{\rm cr}(\omega_{\rm rot})\rangle$,
and is interpreted as an intrinsic state
suitably describing the high-spin state at the spin value
$I\,\hbar\approx
 \langle\Phi_{\rm cr}(\omega_{\rm rot})|
 J_y|\Phi_{\rm cr}(\omega_{\rm rot})\rangle$.
In this way the effects of the collective rotation on the mean-field parameters
like the deformations and the pairing gaps can be well taken into account.

The angular momentum projection is a fully microscopic method to calculate
a sequence of quantum states belonging to one rotational band.
Although we can obtain a whole sequence of the band
by the projection from one intrinsic state,
the resultant spectrum is not always very accurate
especially for the high-spin part; for example,
the calculated moment of inertia is almost constant or even decreases
as a function of spin,
while the observed inertia markedly increases at high-spin states
in most of the realistic cases.
This is due to the fact that the rotational effect, e.g.,
the Coriolis anti-pairing effect, is not taken into account
when only one intrinsic state is employed in the projection.
However, if we intend to combine the cranking model and
the angular momentum projection method,
we encounter a puzzling situation;
namely, we have many spectra depending on the cranking frequencies,
each of which is obtained as a result of angular momentum projection from
one mean-field state associated with each different frequency.
It may be considered that the best way is to choose
the lowest energy state for each spin value; i.e.,
the idea of the variation after projection regarding the cranking frequency
as a variational parameter.
%In fact the cranking model can be derived
%by the method of variation after angular momentum projection~\cite{RS80}.
It has been shown, however, that this does not necessarily work
in practice~\cite{ZSD07};
if the rotational frequency is treated as a variational parameter
at each spin value, it happens that the resultant spectrum does not compose
a regular rotational pattern.
In Ref.~\cite{ZSD07} it is recommended to calculate the spin $I$ state
by the projection from the cranked state with the frequency
$\omega_{\rm rot}$ that gives $ \langle\Phi_{\rm cr}(\omega_{\rm rot})|
 J_y|\Phi_{\rm cr}(\omega_{\rm rot})\rangle \approx I\,\hbar$.
However, this procedure is not efficient because the angular momentum projection
should be performed for each spin state from the intrinsic state with
each different cranking frequency, and there is still no guarantee
that a regular rotational sequence is obtained.

%%%
Thus, how to compromise the cranking model and the angular momentum
projection method is a non-trivial problem.
%%%
In the present work, we attack the problem by a different approach:
The resultant state for each spin value is calculated by
superposing the multiple states obtained by
projection from the cranked mean-field states
with various rotational frequencies,
i.e., in the sense of the configuration mixing or
the generator coordinate method (GCM).
The original idea goes back to the work by Peierls-Thouless~\cite{PT62}.
It will be shown that the resultant spectrum does not essentially
depend on the chosen set of rotational frequencies as long as the number
of them is sufficient; the necessary number is in fact rather small,
something like four to five.  Thus the accurate rotational band
can be efficiently calculated by the configuration mixing, or the GCM,
of a rather small number of angular momentum projected states;
we call this procedure as
``angular momentum projected multi-cranked configuration mixing''.
After a brief explanation of the formulation of the actual procedure
in Sec.~\ref{sec:multi},
we will show three examples of calculation in Sec.~\ref{sec:results}.
Sec.~\ref{sec:summary} is devoted to summary and discussion.

%------------------------------------------------------------------------------
\section{Multi-cranked configuration mixing}
\label{sec:multi}

The method of calculation we adopt is a version of
the angular momentum projected configuration mixing, or the projected GCM,
where the cranking frequency $\omega_{\rm rot}$ is considered
as a generator coordinate.
Thus the resultant state with spin $I$ is obtained in the form
\begin{equation}
 |\Psi^I_{M,\alpha}\rangle = \sum_{Kn} g^I_{Kn,\alpha}\,
 \hat P^I_{MK}|\Phi_n^{}\rangle,
\label{eq:proj}
\end{equation}
where the operator $\hat P^I_{MK}$ is the angular momentum projector
and the amplitude $g^I_{Kn,\alpha}$ is determined by the so-called
Hill-Wheeler equation,
\begin{equation}
 \sum_{K'n'}{\cal H}^I_{Kn,K'n'}\ g^I_{K'n',\alpha} =
 E^I_\alpha\,
 \sum_{K'n'}{\cal N}^I_{Kn,K'n'}\ g^I_{K'n',\alpha},
\label{eq:HW}
\end{equation}
with definitions of the Hamiltonian and norm kernels,
\begin{equation}
 \left\{ \begin{array}{c}
   {\cal H}^I_{Kn,K'n'} \\ {\cal N}^I_{Kn,K'n'} \end{array}
 \right\} = \langle \Phi_n |
 \left\{ \begin{array}{c}
   H \\ 1 \end{array}
 \right\} \hat{P}_{KK'}^I | \Phi_{n'} \rangle,
\label{eq:kernels}
\end{equation}
see e.g. Ref.~\cite{RS80} for details.
A set of the intrinsic states, $|\Phi_n^{}\rangle$,
from which the projection is performed, is chosen to be
composed of mean-field states, i.e., Hartree-Fock-Bogoliubov (HFB) states,
which are obtained by the selfconsistent cranking model
based on the Routhian $H'$ in Eq.~(\ref{eq:routh})
with a suitably chosen set of rotational frequencies,
($\omega^{(n)}_{\rm rot}$; $n=1,2,\cdots,n_{\rm max}$), i.e.
$|\Phi_n^{}\rangle=|\Phi_{\rm cr}(\omega^{(n)}_{\rm rot})\rangle$. 
Then the state in Eq.~(\ref{eq:proj}) is nothing else but
the discrete version of
\begin{equation}
 |\Psi^I_{M,\alpha}\rangle =
 \int d\omega_{\rm rot}\sum_{K} g^I_{K,\alpha}(\omega_{\rm rot})\,
 \hat P^I_{MK}|\Phi_{\rm cr}(\omega_{\rm rot})\rangle,
\label{eq:PTanz}
\end{equation}
which was originally considered by Peierls-Thouless~\cite{PT62}.

Instead of the amplitude $g^I_{Kn,\alpha}$ in Eq.~(\ref{eq:proj}),
the properly normalized amplitude~\cite{RS80},
\begin{equation}
f^I_{Kn,\alpha}=\sum_{K'n'}
 \bigl(\sqrt{{\cal N}^I}\,\bigr)_{Kn,K'n'}\,g^I_{K'n',\alpha},
\label{eq:normf}
\end{equation}
should be considered in order to study the configuration mixing.
For example, the probability of the $n$-th HFB state
in the eigenstate $|\Psi^I_{M,\alpha}\rangle$ is given by
\begin{equation}
p^I_{\alpha}(n)=\sum_{K}|f^I_{Kn,\alpha}|^2,
\label{eq:probf1}
\end{equation}
or for Eq.~(\ref{eq:PTanz}),
\begin{equation}
p^I_{\alpha}(\omega_{\rm rot})=\sum_{K}|f^I_{K,\alpha}(\omega_{\rm rot})|^2.
\label{eq:probf2}
\end{equation}

In the present work we only consider the angular momentum projection
for simplicity,
but the number and parity projections should be incorporated if necessary.
However, the conservation of neutron and/or proton number is important
for the configuration mixing calculation, and we use the approximate way to 
take it into account~\cite{BDF89} when the system is in the superfluid phase;
i.e., the Hamiltonian kernel in Eq.~(\ref{eq:HW}) is replaced by
\begin{equation}
 {\cal H}^I_{Kn,K'n'}\ \rightarrow\ {\cal H}^I_{Kn,K'n'}
 -\lambda_\nu\,\Delta N^I_{Kn,K'n'}
 -\lambda_\pi\,\Delta Z^I_{Kn,K'n'},
\label{eq:cHkernel}
\end{equation}
with
\begin{equation}
\Delta N^I_{Kn,K'n'} =
\langle \Phi_n | (\hat{N}-N_0)\hat{P}_{KK'}^I | \Phi_{n'} \rangle,\quad
\Delta Z^I_{Kn,K'n'} =
\langle \Phi_n | (\hat{Z}-Z_0)\hat{P}_{KK'}^I | \Phi_{n'} \rangle,
\label{eq:numKNZ}
\end{equation}
where the operator $\hat{N}$ ($\hat{Z}$) is the neutron (proton) number operator
and $N_0$ ($Z_0$) is the number to be fixed.
The parameters $\lambda_\nu$ and $\lambda_\pi$ in Eq.~(\ref{eq:cHkernel}) are
taken to be the calculated neutron and proton chemical potentials
of the first HFB state $|\Phi_{n=1}^{}\rangle$;
this choice is enough for the present purpose.

In the following section we will show examples of calculation
performed by the procedure of the angular momentum projected
multi-cranked configuration mixing described above.
We have recently developed an efficient method for
the general quantum-number projection and GCM~\cite{TS12},
and it has been applied
to the study of the nuclear tetrahedral deformation~\cite{TSD13,TSD15}.
The method is fully utilized also in the present work.
As for the effective interaction we employ the finite range
Gogny interaction~\cite{DeGo80} with the D1S parameter set~\cite{D1S}.
The standard method of the harmonic-oscillator basis expansion
is used for the HFB as well as the projection calculations.
The calculational details related to the Gogny interaction
are the same as those explained in Ref.~\cite{TSD15} except for one point:
The Coulomb exchange contribution is treated in the Slater approximation
although we are able to perform the exact calculation.
This is because the selfconsistently calculated pairing correlation
for protons quite often vanishes (or becomes very small)
in the normal deformed nuclei in the rare earth region
due to the fact that the Coulomb antipairing effect is too strong
in the exact treatment~\cite{AER01}.  This deficiency would be avoided
if the variation after the number projection is performed~\cite{AER02}.
The proton pairing correlation is important to describe the collective rotation
and we use the HFB calculation in the present work.
Therefore the Slater approximation is the simplest way to avoid the deficiency.

%------------------------------------------------------------------------------
\section{Examples of calculation}
\label{sec:results}

In the present work we consider three examples of application;
the ground state band of a typical rare-earth nucleus $^{164}$Er,
the ground state band of a very neutron-rich nucleus $^{40}$Mg,
and the superdeformed band of a representative nucleus $^{152}$Dy.
It has been found that the non-cranked HFB calculation gives
axially symmetric deformation in these three examples.
We will discuss the $\lambda$-pole deformation parameter
of the calculated mean-field defined by
\begin{equation}
\beta_{\lambda}\equiv
 \frac{4\pi}{3}\, \frac{\displaystyle
 \biggl\langle \sum_{i=1}^{A}(r^\lambda Y_{\lambda 0})^{}_i \biggr\rangle}
 {A\, \bar{R}^\lambda},\qquad\mbox{with}\quad \bar{R}=
 \sqrt{ \frac{5}{3A}\biggl\langle \sum_{i=1}^{A}r_i^2 \biggr\rangle},
 \label{eq:defparm}
\end{equation}
and the average pairing gap by~\cite{BRRM00}
\begin{equation}
\bar{\Delta}\equiv
\frac{\displaystyle -\sum_{a>b}\Delta_{ab}\kappa^*_{ab}}
 {\displaystyle \sum_{a>b}\kappa^*_{ab}},\qquad\mbox{with}\quad 
  \Delta_{ab}= \sum_{c>d}\bar{v}_{ab,cd}\,\kappa_{cd},
 \label{eq:gapparm}
\end{equation}
where the quantity $\kappa_{ab}$ is the abnormal density matrix
(the pairing tensor) and $\Delta_{ab}$ is the matrix element
of the pairing potential with the anti-symmetrized matrix element
$\bar{v}_{ab,cd}$ of the two-body interaction~\cite{RS80}.

As for the oscillator basis expansion the frequency
$\hbar\omega=41/A^{-1/3}$~MeV is employed
and all the basis states with the oscillator quantum numbers $(n_x,n_y,n_z)$
satisfying $n_x+n_y+n_z \le N_{\rm osc}^{\rm max}$ are retained.
We mainly use $N_{\rm osc}^{\rm max}=12$; however, smaller values are
also adopted for some calculations which require heavy numerical effort.
For the calculation of the angular momentum projector in Eq.~(\ref{eq:proj}),
the Gaussian quadrature over the three Euler angles
$(\alpha,\beta,\gamma)$ is necessary:
Even if the non-cranked HFB calculation gives the axially symmetric solution,
the cranking term in Eq.~(\ref{eq:routh}) breaks
the axial symmetry of the system.
The number of integration mesh points for the angle $\alpha$,
which is the same for the angle $\gamma$,
is taken to be $N_\alpha=N_\gamma=2K_{\rm max}+2$,
and that for the angle $\beta$ to be $N_\beta=2I_{\rm max}+2$,
where $I_{\rm max}$ and $K_{\rm max}$ are the maximum values
of the $I$ and $K$ quantum numbers suitably chosen for each case.

It has been known that the density-dependent term in the Skyrme interaction
causes various numerical problems,
see e.g. Refs.~\cite{DSN07,Rob10} and references therein.
The Gogny interaction has the same density-dependent term
and we encounter the situations that
unphysical spectra are obtained due to the density-dependent term;
for example, the spectra in the well-deformed odd mass nuclei
sometimes do not look like rotational.
The reason why we judge that it is due to the density-dependent term is
that quite nice rotational spectra are recovered if we exclude it,
although then the absolute energy and the moment of inertia are
different from what they should be.
It seems to us that the time-odd components of the wave function
induced by the cranking term tend to increase the rate of occurrence
of this problem particularly for odd and odd-odd nuclei.
However, we only consider even-even nuclei in the present work
and the considered cranking frequency is not very high.
Any problems related to the density-dependent term did not seem
to occur in the following examples.

Another remark is related to the so-called norm cut-off;
the states whose norm eigenvalues are smaller than a certain value
are excluded when solving the Hill-Wheeler equation~(\ref{eq:HW}).
The value for cut-off should be small enough
not to miss important contributions,
while it should not be too small in order to avoid the numerical difficulty
in the GCM related to the vanishing norm states~\cite{RS80}.
We generally try to find a proper value as small as possible.
It is, however, necessary to choose a suitable value
for each calculation depending on the situation.
This is because we use the same norm cut-off
for all spin members under consideration
in the calculation of a whole rotational band.
It could happen that some small norm states included at some spins,
which is controlled by the value of the norm cut-off,
get excluded at different spins when solving Eq.~(\ref{eq:HW}).
Such a change of included states at different spins may cause
discontinuities of the rotational spectrum.   Therefore we should adjust
value of the norm cut-off so that any discontinuities would not happen;
we are able to find such a proper value for the norm cut-off.

%###################################################
\subsection{Ground state band of \,$^{\it 164}$\hspace*{-2.5pt}Er}
\label{sec:Er-gr}

The first example is the ground state rotational band of a nucleus $^{164}$Er,
where the high-spin states up to $I^\pi=22^+$ are measured~\cite{Er164gr}.
We use the oscillator basis with $N_{\rm osc}^{\rm max}=12$.
The selfconsistently calculated values of the mean-field parameters
of the ground state in this nucleus are $\beta_2=0.311$
for the quadrupole deformation,
which roughly corresponds to the experimentally deduced value~\cite{LVH70},
and $\bar{\Delta}=0.874,\,0.906$~MeV
for the neutron and proton average pairing gaps, respectively,
which are about 10\%--13\% smaller than the even-odd mass differences.
In this case the system is in the superfluid phase, and we apply
the quasiparticle basis cut-off at the level of $10^{-6}$,
i.e., only the canonical basis states whose occupation probabilities
are larger than $10^{-6}$ are considered, see Ref.~\cite{TS12} for details.
As for the projection calculation the maximum values of
the $I$ and $K$ quantum numbers are taken to be
$I_{\rm max}=22$ and $K_{\rm max}=16$ and therefore the numbers of integration
mesh points are $N_\alpha=N_\gamma=34$ and $N_\beta=46$ for the Euler angles.

\begin{figure}[!htb]
\begin{center}
\includegraphics[width=70mm]{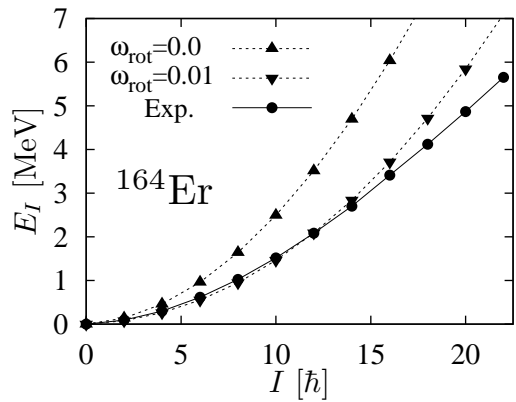}
\vspace*{-4mm}
\caption{
Excitation energy spectra of the ground state rotational band in $^{164}$Er
calculated by the projection method from
the non-cranked HFB state (the label ``$\omega_{\rm rot}=0.0$'') and
from the cranked HFB state with $\hbar\omega_{\rm rot}=0.01$~MeV
 (``$\omega_{\rm rot}=0.01$'').
Experimentally measured spectrum (``Exp.'') is also included.
The $0^+$ energy is taken as the energy origin in each spectrum.
The oscillator basis with $N_{\rm osc}^{\rm max}=12$ is used.
}
\label{fig:Erprj1}
\end{center}
\end{figure}

First we show the calculated excitation spectrum obtained by
the angular momentum projection from the non-cranked HFB state
(i.e. $\omega_{\rm rot}=0$) in comparison with the experimental data
in Fig.~\ref{fig:Erprj1}. The energy gain of the ground state
by the angular momentum projection is 2.97~MeV.
The value of norm cut-off can be taken to be $10^{-11}$ in this calculation.
As it is clearly seen in the figure, the projected rotational energies
from the non-cranked state are systematically larger than the experimental ones.
We also include the result of projection from the cranked HFB state
with a small cranking frequency $\hbar\omega_{\rm rot}=0.01$~MeV
in Fig.~\ref{fig:Erprj1}.
The curvature of the spectrum as a function of spin decreases considerably
as a result of including the time-odd components in the wave function
by the cranking model,
and the agreement with the experimental data is much better.
Namely the calculated moment of inertia
is enlarged by the small cranking term nearly by a factor of two.
This result clearly shows the importance of the cranking procedure
for the angular momentum projection calculation,
which was emphasized in Refs.~\cite{TS12,TSD13}.
However, the rotational energies are overestimated
at the higher spin states, $I \ge 16$,
although they are slightly underestimated at the lower spin states, $I \le 12$:
The experimental moment of inertia increases at higher-spin states,
while the calculated inertias are almost constant as long as the projection
is performed from only one intrinsic HFB state,
see also Figs.~\ref{fig:ErMoIm} and~\ref{fig:ErMoIm5}.
Thus it is necessary to improve the calculation
by the multi-cranked configuration mixing.

\begin{figure}[!htb]
\begin{center}
\includegraphics[width=70mm]{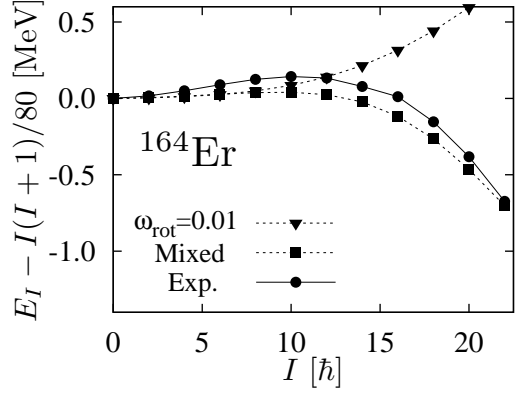}
\vspace*{-4mm}
\caption{
Excitation energy spectra subtracting the reference rotational energy,
$I(I+1)/80$~MeV, for $^{164}$Er.
The result of the simple projection from one cranked HFB state
with $\hbar\omega_{\rm rot}=0.01$~MeV as well as that
of the projected multi-cranked configuration mixing (the label ``Mixed'')
explained in the text are included in addition to the experimental data.
}
\label{fig:Erprj2}
\end{center}
\end{figure}

\begin{figure}[!htb]
\begin{center}
\includegraphics[width=80mm]{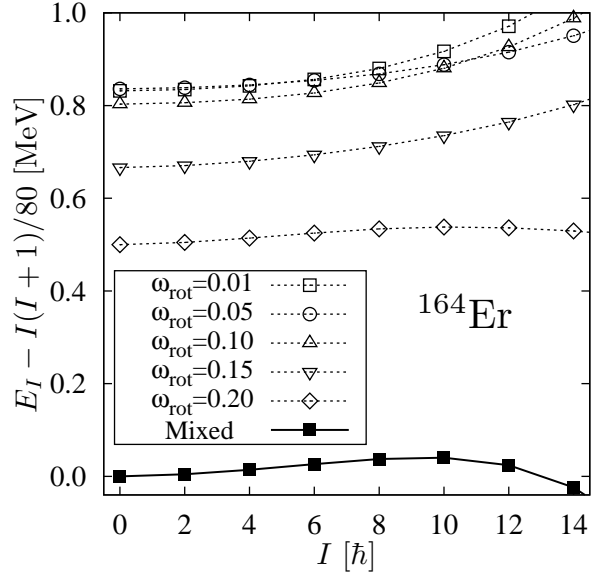}
\vspace*{-4mm}
\caption{
Energy spectra for $^{164}$Er obtained by
the simple projection from one intrinsic HFB state
with five values of the cranking frequencies,
$\hbar\omega_{\rm rot}=0.01,\,0.05,\,0.10,\,0.15,\,0.20$~MeV,
compared with the result of the projected configuration mixing
employing those five HFB states.
The energy origin is taken as the energy of the $I=0$ state of
the configuration mixing calculation, and
the same reference rotational energy is subtracted as in Fig.~\ref{fig:Erprj2}.
}
\label{fig:Ermix5}
\end{center}
\end{figure}

We show in Fig.~\ref{fig:Erprj2} the results of 
the projected multi-cranked configuration mixing
and of the simple projection from only one cranked HFB state
with $\hbar\omega_{\rm rot}=0.01$~MeV in comparison with the experimental data.
The reference rotational energy
$I(I+1)\hbar^2/(2{\cal J})$ with ${\cal J}=40~[\hbar^2/$MeV]
is subtracted to display the detailed behaviors more clearly.
A set of five almost equidistant cranking frequencies,
$(\hbar\omega_{\rm rot}^{(n)},\,n=1:5)
=(0.01,\,0.05,\,0.10,\,0.15,\,0.20)$~MeV, is adopted.
The agreement with the experimental data is much better;
the difference of excitation energy is less than 130~keV
in the whole spin range, $0 \le I \le 22$,
if the configuration mixing is performed.
In order to display the effect of the configuration mixing,
the five spectra obtained by the projection from one intrinsic HFB state
with the five different cranking frequencies are shown in Fig.~\ref{fig:Ermix5}
in addition to the result of the projected configuration mixing.
The energy origin is chosen to be the $0^+$ energy
of the final configuration mixing and
the reference rotational energy is subtracted
as in the same way as in Fig.~\ref{fig:Erprj2}.
The resultant rotational spectra are similar as long as
one cranked HFB state is used, while the absolute $0^+$ energies are smaller
when projected from the HFB state with higher cranking frequencies.
The energy gain of the configuration mixing from the simple projection
with the non-cranked HFB state is about 0.83~MeV and
the total energy gain by the projected configuration mixing
from the HFB energy is 3.81~MeV.
It is quite interesting to mention that
the excited $K^\pi=0^+$ rotational band of the Hill-Wheeler equation
as a result of the configuration mixing has about 6~MeV higher energy
than the ground state. This is consistent to the fact that
there is only one rotational band associated with the ground state,
and the five rotational sequences obtained by the simple projection
from each configuration are not totally independent;
in fact the overlaps between these five HFB states are rather large.

\begin{figure}[!htb]
\begin{center}
\includegraphics[width=70mm]{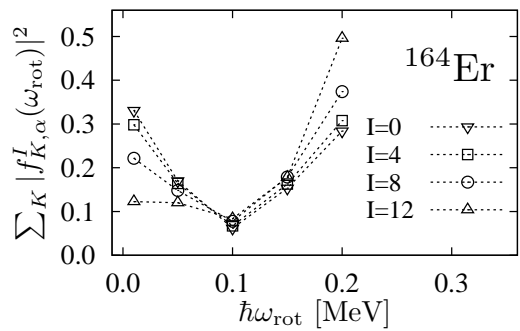}
\vspace*{-4mm}
\caption{
Probability distribution (Eq.~(\ref{eq:probf2}))
over the five HFB configurations with different cranking frequencies
for the spin $I$ member of the ground state rotational band of $^{164}$Er
obtained by the projected configuration mixing.
}
\label{fig:Erprob}
\end{center}
\end{figure}

The probabilities defined in Eq.~(\ref{eq:probf2}) of the five configurations
associated with $\hbar\omega_{\rm rot}=0.01,\,0.05,\,0.10,\,0.15,\,0.20$~MeV
are depicted in Fig.~\ref{fig:Erprob}.
The distributions shown in Fig.~\ref{fig:Erprob}
are rather different from what are expected, i.e.,
the probabilities have peaks at the cranking frequencies
corresponding to $ \langle\Phi_{\rm cr}(\omega_{\rm rot})|
 J_y|\Phi_{\rm cr}(\omega_{\rm rot})\rangle \approx I\,\hbar$.
In fact the distributions spread over five configurations with
the middle one ($\hbar\omega_{\rm rot}=0.10$ MeV)
having always the lowest probability.
In fact the distributions for all the $I=0,\,4,\,8,\,12$ members are
rather similar and the main differences are the probabilities
of the first ($\hbar\omega_{\rm rot}=0.01$ MeV) and
the last ($\hbar\omega_{\rm rot}=0.20$ MeV) configurations.
It should be pointed out that the rotational sequence projected from
the last configuration is lowest in energy, see Fig.~\ref{fig:Ermix5},
and therefore its probabilities are relatively large
for all the $I=0,\,4,\,8,\,12$ members, which may be specific in this example.

\begin{figure}[!htb]
\begin{center}
\includegraphics[width=70mm]{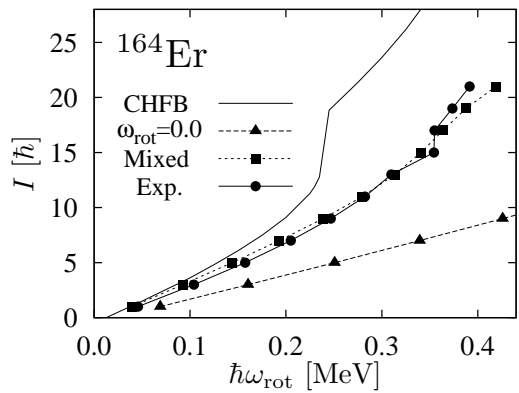}
\vspace*{-4mm}
\caption{
Angular momentum versus rotational frequency for $^{164}$Er
obtained by various calculations;
the cranked HFB (the label ``CHFB''),
the simple projection from the non-cranked HFB state
(``$\omega_{\rm rot}=0.0$''),
and the projected configuration mixing (``Mixed'')
in comparison with the experimental data (``Exp.'').
}
\label{fig:ErIom}
\end{center}
\end{figure}

Figure~\ref{fig:ErIom} depicts the angular momenta as functions
of the cranking frequency for various calculations
in comparison with the experimental data.
For the cranked HFB calculation the angular momentum is calculated by
$I(\omega_{\rm rot})\,\hbar \equiv\langle\Phi_{\rm cr}(\omega_{\rm rot})|
 J_y|\Phi_{\rm cr}(\omega_{\rm rot})\rangle - \frac{1}{2}\hbar$
as a function of the rotational frequency $\omega_{\rm rot}$,
where the subtraction of $\frac{1}{2}\hbar$ is the semiclassical correction,
see e.g. Ref.~\cite{BF79}.  For the results of
projection calculation either from the non-cranked HFB state
or with the configuration mixing, the rotational frequency is
calculated by the finite difference,
$\hbar\omega_{\rm rot}(I)\equiv(E(I+1)-E(I-1))/2$,
where $E(I)$ is the energy at spin~$I$.
One can see that the slopes at lower spins are quite different
for various calculations
and again the agreement of the configuration mixing calculation with
the experimental data is best, while the slope of the cranked HFB is 
larger and that of the simple projection from the non-cranked HFB state
($\hbar\omega_{\rm rot}=0.0$) is considerably smaller
than the experimental data. It should be noticed that
the abrupt upbend of the angular momentum is observed
in the cranked HFB calculation at $\hbar\omega_{\rm rot}\approx 0.25$~MeV.
This is caused by the angular-momentum-alignment of
the lowest two quasi-neutrons along the rotational axis,
which leads to the crossing
between the ground state band (g-band) and the Stockholm band (s-band).
This band-crossing has been known for many years as
the origin of the backbending phenomenon first observed in Ref.~\cite{JRH72},
see e.g. discussions in Refs.~\cite{VDS83,GHH86}.
The effect of this band-crossing is reflected as an irregularity at $I=15$
in the experimental data in Fig.~\ref{fig:ErIom}, while there is
no such effect for the projected calculations because
the crossing with the s-band configuration is not considered.

It should be mentioned that the selfconsistent cranked HFB calculation
shows the characteristic changes of
the mean-field parameters as functions of the rotational frequency.
For example, the quadrupole deformation parameter $\beta_2$
slightly increases until about $\hbar\omega_{\rm rot}\approx 0.2$~MeV
and starts to decreases, although the total amount of change is less
than 10\% in the depicted range of frequency in Fig.~\ref{fig:ErIom}.
The triaxiality parameter~$\gamma$ gradually increases from zero
at zero frequency up to a small value about $\gamma\approx +7^\circ$
at $\hbar\omega_{\rm rot}\approx 0.24$~MeV.
The average pairing gaps for both neutrons and protons gradually decrease
and that of neutrons suddenly drops after the g-s crossing
because of the blocking effect.
These behaviors of the mean-field parameters have been well-known for
well deformed nuclei in the rare-earth region by
similar selfconsistent cranking calculation but
with simple schematic interactions,
see e.g. Ref.~\cite{EMR80} for this nucleus.
In fact the results of the present cranked HFB calculation are very similar
to those of Ref.~\cite{EMR80} (note that the cranking axis is the $y$-axis
in the present work while it is the $x$-axis in Ref.~\cite{EMR80}) except for
the fact that the neutron pairing gap vanishes after the crossing frequency
$\hbar\omega_{\rm rot}\approx 0.24$~MeV in the present calculation.
This is the main reason why we do not investigate the s-band
in the present work: The calculated moment of inertia for the s-band
is overestimated due to the vanishing neutron pairing correlation.
We have used the maximum cranking frequency,
$\hbar\omega_{\rm rot}=0.20$~MeV, for the configuration mixing calculation
of the g-band because of a similar reason; the effects of the s-band
configuration should be excluded for describing the g-band.

The present cranked HFB calculation in Fig.~\ref{fig:ErIom} does not show
the backbending behavior like in Ref.~\cite{EMR80},
because the minimum-search with the proper treatment of
the angular momentum constraint~\cite{RS80}
is not performed in the backbending region.
The proper treatment of the band-crossing phenomenon
within the projected configuration mixing approach is now under investigation.

\begin{figure}[!htb]
\begin{center}
\includegraphics[width=70mm]{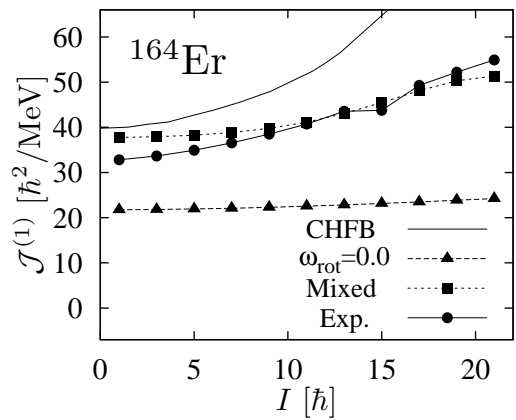}
\vspace*{-4mm}
\caption{
Moments of inertia versus spin value for $^{164}$Er obtained by
various calculations in comparison with the experimental data
like in Fig.~\ref{fig:ErIom};
see the text for the precise definition of
the first moment of inertia ${\cal J}^{(1)}$.
}
\label{fig:ErMoIm}
\end{center}
\end{figure}

\begin{figure}[!htb]
\begin{center}
\includegraphics[width=80mm]{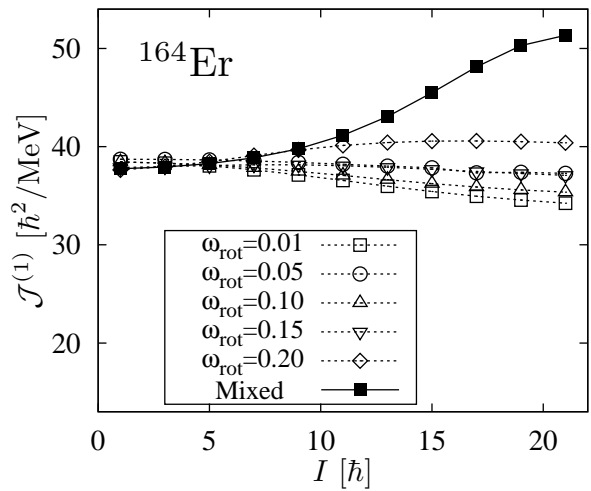}
\vspace*{-4mm}
\caption{
Moments of inertia versus spin value for $^{164}$Er obtained by
the simple projection from one intrinsic HFB state
with five values of the cranking frequency,
$\hbar\omega_{\rm rot}=0.01,0.05,0.10,0.15,0.20$~MeV,
compared with the result of the projected configuration mixing
employing those five HFB states.
The ordinate scale is enlarged from that of Fig.~\ref{fig:ErMoIm}.
}
\label{fig:ErMoIm5}
\end{center}
\end{figure}

In order to study the property of the rotational band more closely,
we depict the result of
the so-called first (or kinematic) moment of inertia ${\cal J}^{(1)}$
for various calculations and the experimental data
in Figs.~\ref{fig:ErMoIm} and~\ref{fig:ErMoIm5}.
It is defined for the cranked HFB calculation by
${\cal J}^{(1)}(\omega_{\rm rot})\equiv \langle\Phi_{\rm cr}(\omega_{\rm rot})|
 J_y|\Phi_{\rm cr}(\omega_{\rm rot})\rangle/\omega_{\rm rot}$,
which is plotted as a function of
$I\,\hbar =\langle\Phi_{\rm cr}(\omega_{\rm rot})|
 J_y|\Phi_{\rm cr}(\omega_{\rm rot})\rangle - \frac{1}{2}\hbar$,
and for the projection calculations and the experimental data
by ${\cal J}^{(1)}(I)\equiv(2I+1)\hbar^2/(E(I+1)-E(I-1))$.
As it is clearly seen in Fig.~\ref{fig:ErMoIm},
the calculated moment of inertia obtained
by the simple projection from the non-cranked HFB state
is considerably underestimated:
Its value is only 66\% of the experimental value at the lowest spin,
while the results of the cranked HFB and of the simple projection from
one cranked HFB state with the finite frequencies
$\hbar\omega_{\rm rot}\le 0.20$~MeV are about 15\% larger than
the experimental value at the lowest spin,
which may be due to the fact that the calculated pairing gaps are 
about 10\% smaller than the even-odd mass differences.
Compared to this, the result of the projected configuration mixing gives
slightly better agreement with the experimental data.
It is interesting to notice that all the results of the simple projection
from one cranked HFB state associated with different frequencies
are rather similar at the lowest spin, see Fig.~\ref{fig:ErMoIm5}.
As for the spin-dependence of the moment of inertia, only the results
of the cranked HFB and of the projected configuration mixing increase
as functions of spin similarly to the experimental moment of inertia.
The increase of calculated inertia by the cranked HFB method
is mainly caused by the gradual reduction of pairing correlations,
i.e., the so-called Coriolis anti-pairing effect.
Those of the simple projection from one HFB state with
various cranking frequencies (including zero-frequency) are rather constant,
or even slightly decreasing for the cases of the finite frequencies,
$\hbar\omega_{\rm rot}=0.01,0.15,0.10,0.15$~MeV, and are completely
different from the experimental spin-dependence.
Thus the configuration mixing is important
in order for the projected result of moment of inertia to increase,
although the amount of increase is not enough compared with
the behavior of the experimental data.

\begin{figure}[!htb]
\begin{center}
\includegraphics[width=155mm]{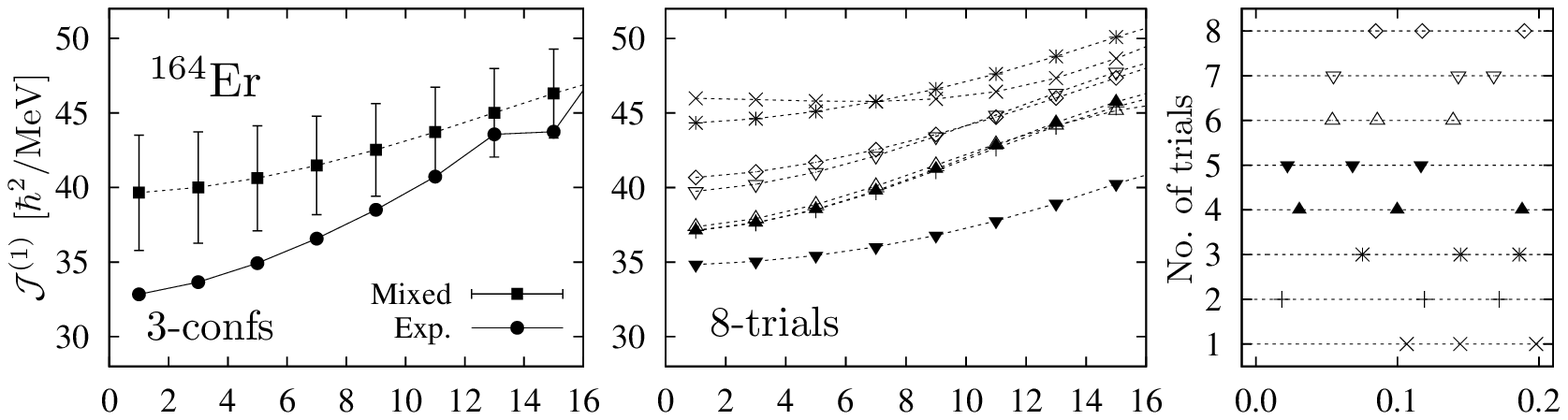}
\includegraphics[width=155mm]{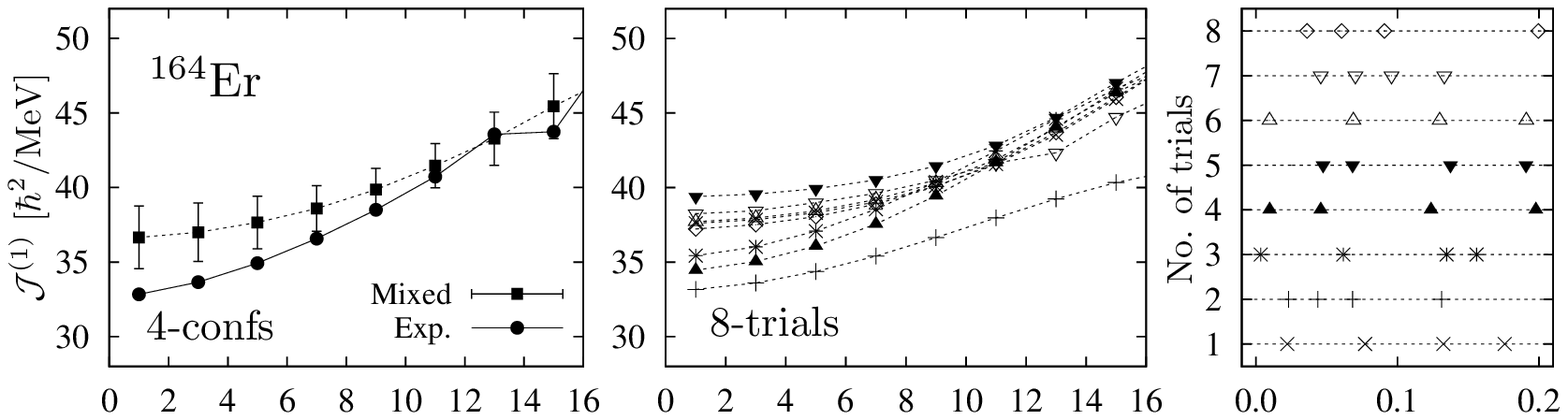}
\includegraphics[width=155mm]{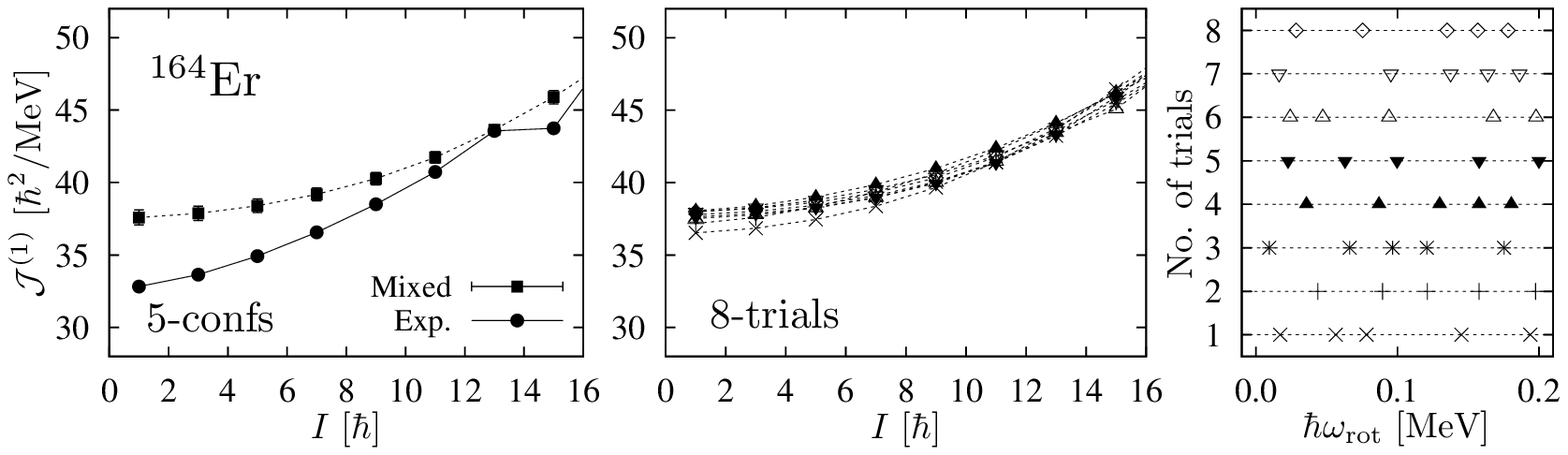}
\vspace*{-4mm}
\caption{
Analysis of moments of inertia for $^{164}$Er calculated by
the projected configuration mixing employing cranked HFB states
with randomly chosen cranking frequencies.
The smaller model space $N_{\rm osc}^{\rm max}=10$ is used in this calculation.
Three sets of three panels in the first, second, and third raw show
the results of calculation with mixing three, four, and five configurations,
respectively, in comparison with the experimental data.
In each set the left panel depicts the average moment of inertia
with the error bars indicating the standard deviation for eight trials of
randomly chosen frequencies, which are shown in the right panel as crosses,
and the middle panel displays the moments of inertia calculated
in those eight trials.
The results of eight trials are distinguished by different symbols.
}
\label{fig:ErRmoi}
\end{center}
\end{figure}

We have used a set of almost equidistant cranking frequencies
to generate the cranked HFB states for the configuration mixing.
There are, however, infinitely many possible choices.
It is interesting to see how the result of configuration mixing
depends on the chosen set of cranking frequencies,
which is analyzed in Fig.~\ref{fig:ErRmoi} by making use of random numbers.
In this calculation the smaller basis size with $N_{\rm osc}^{\rm max}=10$
is used to reduce the numerical task.
We have randomly generated various sets of cranking frequencies
with the conditions that the frequency satisfies,
$0 < \hbar\omega_{\rm rot} < 0.20$~MeV,
to avoid the effect of the g-s band crossing,
and the difference of two nearest frequencies satisfies,
$\hbar\Delta\omega_{\rm rot}>0.02$~MeV,
to avoid too large overlap between the two associated HFB states,
which may cause the vanishing norm problem of the GCM procedure~\cite{RS80}.
It happens that the norm cut-off problem occurs
for these randomly generated frequencies.
Therefore, we had to adjust values of norm cut-off
in the range $10^{-8}-10^{-10}$.
The results of analysis employing three, four, and five configurations
associated with randomly chosen frequencies are shown
in the first, second, and third rows, respectively, in Fig.~\ref{fig:ErRmoi}.
We have performed eight trial calculations in each case,
and the average and standard deviation of the calculated moments of inertia
are plotted in the left panels,
where the  standard deviation is shown as error bars.
The middle panels show all moments of inertia calculated in eight trials,
and the randomly generated sets of cranking frequencies applied in these trials
are displayed in the right panels.
As is seen in Fig.~\ref{fig:ErRmoi},
when the number of configurations is three,
not only the standard deviation is large but also the average value itself
deviates considerably from the one calculated
with larger numbers of configurations.  The difference of calculated moments
of inertia among eight trials is getting smaller
as the number of mixed configurations is increased,
even though the generated sets of frequencies are markedly different.
In fact, if the five configurations are employed the standard deviations
are within the size of symbols,
see the left-bottom panel in Fig.~\ref{fig:ErRmoi}.
Therefore the result of multi-cranked configuration mixing with
the equidistant set of five cranking frequencies gives the almost unique result.
The small difference between the converged result of mixing
the five configurations in Fig.~\ref{fig:ErRmoi}
and that in Fig.~\ref{fig:ErMoIm} or Fig.~\ref{fig:ErMoIm5} is due to
the different model space with $N_{\rm osc}^{\rm max}=10$ and~12 being used.

\begin{figure}[!htb]
\begin{center}
\includegraphics[width=70mm]{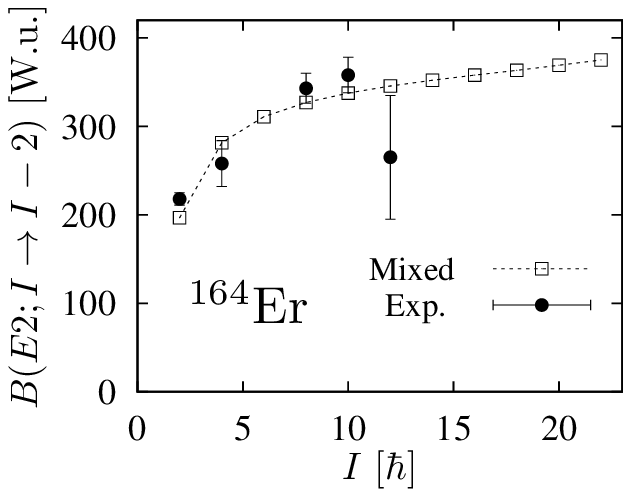}
\vspace*{-4mm}
\caption{
Calculated in-band $B(E2;I\rightarrow I-2)$ values
by the projected configuration mixing in for the ground state band
of $^{164}$Er in comparison with the experimental data.
}
\label{fig:ErBE2}
\end{center}
\end{figure}

\begin{figure}[!htb]
\begin{center}
\includegraphics[width=70mm]{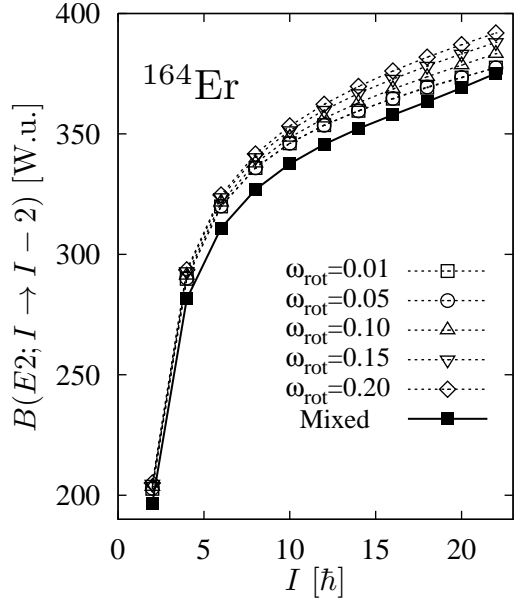}
\vspace*{-4mm}
\caption{
Calculated in-band $B(E2;I\rightarrow I-2)$ values
by the simple projection from one HFB state with different cranking
frequencies in comparison with the result of configuration mixing.
The ordinate scale is enlarged compared with Fig.~\ref{fig:ErBE2}.
}
\label{fig:ErBE2m5}
\end{center}
\end{figure}

At the end of this subsection we briefly discuss the transition probabilities.
We compare the calculated in-band $B(E2)$ values
with experimental data~\cite{Er164BE2} in Fig.~\ref{fig:ErBE2};
a good agreement is achieved.
In order to show the effect of configuration mixing for the $B(E2)$ values,
we show those calculated by the simple projection from one HFB state
with different cranking frequencies in Fig.~\ref{fig:ErBE2m5}.
The calculated $B(E2)$ value is larger when projected from
the HFB state with higher frequency; this can be well understood
because the deformation parameter $\beta_2$ slightly increases
and the triaxiality parameter $\gamma$ is positive and also slightly increases
in this frequency range ($0 \le \hbar\omega_{\rm rot} \le 0.2$ MeV),
both of which lead to the increase of quadrupole moment
around the rotation axis (note that it is the $y$-axis in the present work).
As it is shown in Fig.~\ref{fig:ErBE2m5}, however,
the configuration mixing makes the $B(E2)$ values smaller than
any results of the simple projection from one HFB state
at all spin values considered, although the amount of reduction
is rather small, less than 5\%.
We think that this is due to the fact that the mixing probabilities
rather spread over five configurations as was shown in Fig.~\ref{fig:Erprob},
which reduces the expectation value of the $E2$ operator by decoherence.
As it is well known that the $B(E2)$ in the rotational band
can be nicely fitted by the rotor model~\cite{BM75}:
In fact the quadrupole moment is usually extracted from
the experimental data assuming this model.
The angular momentum projection calculation of the $B(E2)$
for well-deformed nuclei justifies the rotor model, see e.g. Ref.~\cite{HS95}.
Since the deformation parameter of the selfconsistent HFB calculation
almost reproduces the correct quadrupole deformation,
the agreement in Fig.~\ref{fig:ErBE2} may be natural.

%###################################################
\subsection{Ground state band of \,$^{\it 40}$\hspace*{-2.5pt}Mg}
\label{sec:Mg-gr}

We take a nucleus $^{40}$Mg as an example of the deformed unstable nuclei:
It has been predicted, see e.g. Ref.~\cite{TFHB97},
that the $N=28$ Mg isotope is deformed in spite of the fact
that $N=28$ is one of the magic numbers in stable nuclei.   This nucleus
is also predicted to be the drip-line nucleus in Ref.~\cite{TFHB97}.
We use the oscillator basis with $N_{\rm osc}^{\rm max}=12$ for this example.
The selfconsistently calculated value of the quadrupole deformation parameter
is $\beta_2=0.334$, and the pairing correlations
for both neutrons and protons vanish.
The neutron skin develops considerably in this nucleus and
the skin thickness is near 1 fm along the short-axis of deformation.
The calculated root-mean-square radius is 3.63 fm,
which is about 15\% larger than the value given by
the empirical radius $1.2A^{1/3}$ fm.
It should be noted that the use of the calculated radius $\bar{R}$
in the definition~(\ref{eq:defparm}) is important to estimate the deformation; 
the deformation parameter is largely overestimated
if the value $\bar{R}=1.2A^{1/3}$ fm is employed instead.
These basic features of the present result of the Gogny HFB calculation
well correspond to those of Ref.~\cite{TFHB97}, although the calculated
deformation is slightly larger in our calculation.
As for the projection calculation we take $I_{\rm max}=20$ and $K_{\rm max}=12$
and therefore the numbers of integration mesh points are
$N_\alpha=N_\gamma=26$ and $N_\beta=42$ for the Euler angles.
The preliminary report of our calculation for this nucleus
was published in Ref.~\cite{TSFSD14}.

\begin{figure}[!htb]
\begin{center}
\includegraphics[width=70mm]{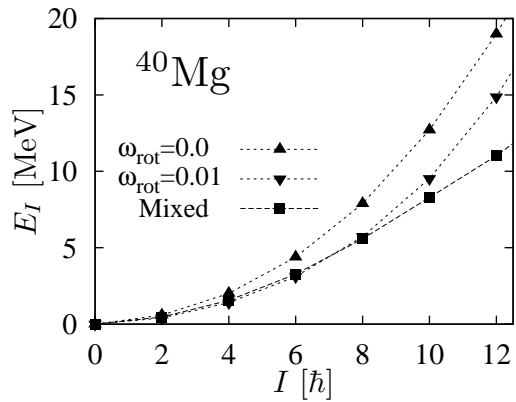}
\vspace*{-4mm}
\caption{
Excitation energy spectrum of the ground state rotational band in $^{40}$Mg
calculated by the projected configuration mixing (the label ``Mixed'')
in comparison with those by the simple projection from
the non-cranked HFB state (``$\omega_{\rm rot}=0.0$'') and
from the cranked HFB state with $\hbar\omega_{\rm rot}=0.01$~MeV
 (``$\omega_{\rm rot}=0.01$'').
The oscillator basis with $N_{\rm osc}^{\rm max}=12$ is used.
}
\label{fig:Mgprj1}
\end{center}
\end{figure}

The excitation spectra calculated by the angular momentum projection method
are shown in Fig.~\ref{fig:Mgprj1}, where are shown
three spectra obtained by
the simple projection from the non-cranked HFB state and
from the cranked HFB state with a small frequency
$\hbar\omega_{\rm rot}=0.01$~MeV,
and by the projected configuration mixing.
A set of four equidistant cranking frequencies,
$(\hbar\omega_{\rm rot}^{(n)},\,n=1:4)$
=$(0.01,\,0.34,\,0.67,\,1.00)$~MeV,
is adopted for calculating the cranked HFB states employed
in the configuration mixing calculation.
There is no experimental data available yet for this nucleus.
The value of norm cut-off is taken to be $10^{-8}$ in this calculation.
The energy gain of the ground state by the simple angular momentum projection
is 3.07~MeV and that by the configuration mixing is 3.47~MeV.
The projected GCM calculation for this nucleus with
the same Gogny D1S interaction was reported in Ref.~\cite{RER02}.
Although the detailed treatment is different from the present calculation,
the results are quite consistent with each other.

\begin{figure}[!htb]
\begin{center}
\includegraphics[width=80mm]{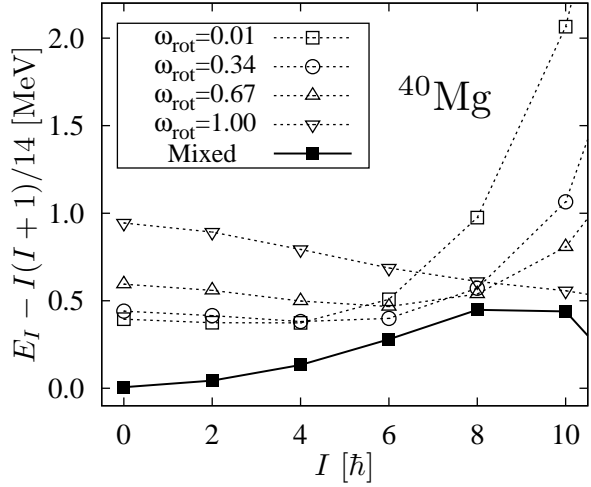}
\vspace*{-4mm}
\caption{
Energy spectra for $^{40}$Mg obtained by
the simple projection from one intrinsic HFB state
with four values of the cranking frequency,
$\hbar\omega_{\rm rot}=0.01,\,0.34,\,0.67,\,1.00$~MeV,
compared with the result of the projected configuration mixing
employing those four HFB states.
The energy origin is taken as the energy of the $I=0$ state of
the configuration mixing calculation, and
the reference rotational energy, $I(I+1)/14$~MeV, is subtracted.
}
\label{fig:Mgmix4}
\end{center}
\end{figure}

\begin{figure}[!htb]
\begin{center}
\includegraphics[width=70mm]{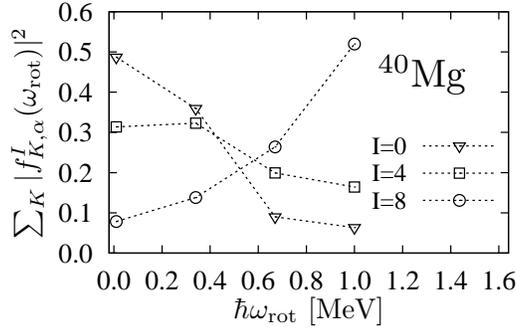}
\vspace*{-4mm}
\caption{
Probability distribution (Eq.~(\ref{eq:probf2}))
over the four HFB configurations with different cranking frequencies
for the spin $I$ member of the ground state rotational band of $^{40}$Mg
obtained by the projected configuration mixing.
}
\label{fig:Mgprob}
\end{center}
\end{figure}

The effect of the configuration mixing is shown in Fig.~\ref{fig:Mgmix4},
where the four spectra obtained by the projection from
one intrinsic HFB state with four cranking frequencies,
$\hbar\omega_{\rm rot}=0.01,\,0.34,\,0.67,\,1.00$~MeV, are depicted
in addition to the result of the projected configuration mixing employing
those cranked HFB states.
The probabilities defined in Eq.~(\ref{eq:probf2})
of these four configurations are also shown in Fig.~\ref{fig:Mgprob}.
The energy origin is chosen to be the $0^+$ energy
of the final configuration mixing and the reference rotational energy,
$I(I+1)/14$~MeV, is subtracted.
The resultant rotational spectra of the projection from one cranked HFB state
are similar at lower spins but considerably different at higher spins.
The absolute $0^+$ energies are larger
when is used the HFB state with higher cranking frequency,
which is opposite to the case of the previous example $^{164}$Er.
The excited $K^\pi=0^+$ rotational band in the configuration mixing
has about 2.7~MeV higher energy than the ground state,
and again it indicates that
there is only one rotational band associated with the ground state.
The probability distributions in Fig.~\ref{fig:Mgprob} seem to be
consistent with what are expected, i.e., the probabilities have peaks
roughly at the cranking frequencies corresponding to
$ \langle\Phi_{\rm cr}(\omega_{\rm rot})|
 J_y|\Phi_{\rm cr}(\omega_{\rm rot})\rangle \approx I\,\hbar$,
see also the following Fig.~\ref{fig:MgIom}.

\begin{figure}[!htb]
\begin{center}
\includegraphics[width=70mm]{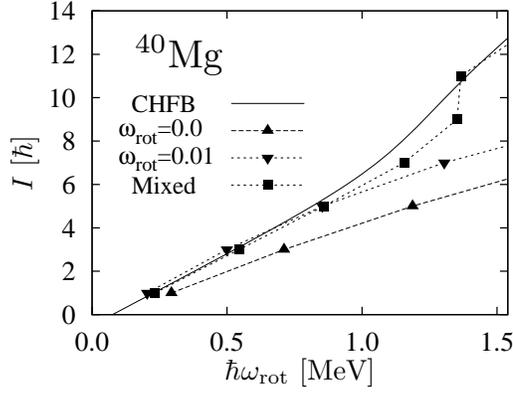}
\vspace*{-4mm}
\caption{
Angular momentum versus rotational frequency for $^{40}$Mg
obtained by various calculations;
the cranked HFB (the label ``CHFB''),
the projection from the non-cranked HFB state
(``$\omega_{\rm rot}=0.0$''),
the projection from the cranked HFB state
with $\hbar\omega_{\rm rot}=0.01$~MeV (``$\omega_{\rm rot}=0.01$''),
and the projected configuration mixing (``Mixed'').
}
\label{fig:MgIom}
\end{center}
\end{figure}

Figure~\ref{fig:MgIom} depicts the angular momenta as functions
of the cranking frequency for various calculations.
One can see that the slope of the result of the projection
from the non-cranked HFB state is smaller than the others,
while the three results of the cranked HFB,
of the simple projection from the cranked HFB state with
$\hbar\omega_{\rm rot}=0.01$~MeV, and of the projected configuration mixing
are rather similar at lower frequencies.  However, at higher frequencies,
$\hbar\omega_{\rm rot}\approx 1.3$~MeV, the upbending behavior can be seen
for the projected configuration mixing calculation.  We do not understand
the precise reason but it seems that the rotational band at higher spins,
$I \ge 12$, changes its nature.  This behavior does not disappear even if
the cranked HFB states with higher frequencies, $\hbar\omega_{\rm rot}>1.0$~MeV,
are included in the configuration mixing calculation.
Taking a closer look at the result of the cranked HFB in Fig.~\ref{fig:MgIom},
the slope gradually increases at the higher frequency,
$\hbar\omega_{\rm rot}\gtrsim 1.2$~MeV,
which is caused by the gradual rotational alignment.
The last occupied neutron orbit $[321\,\frac{1}{2}]$ is bound but
its binding energy, $1$ MeV at $\omega_{\rm rot}=0$, decreases
to about 150 keV at higher frequency $\hbar\omega_{\rm rot}\approx 1.4$~MeV,
where this orbit strongly interacts with a discretized continuum state;
note that the pairing correlations vanish in this nucleus.
Therefore we only consider the rotational spectrum in the spin range, $I<12$.

\begin{figure}[!htb]
\begin{center}
\includegraphics[width=70mm]{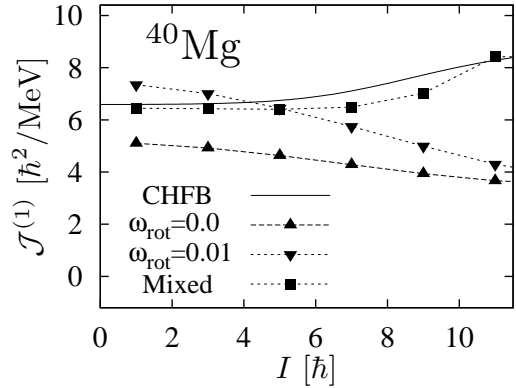}
\vspace*{-4mm}
\caption{
Moments of inertia versus spin value for $^{40}$Mg obtained by
various calculations like in Fig.~\ref{fig:MgIom}.
}
\label{fig:MgMoIm}
\end{center}
\end{figure}

\begin{figure}[!htb]
\begin{center}
\includegraphics[width=80mm]{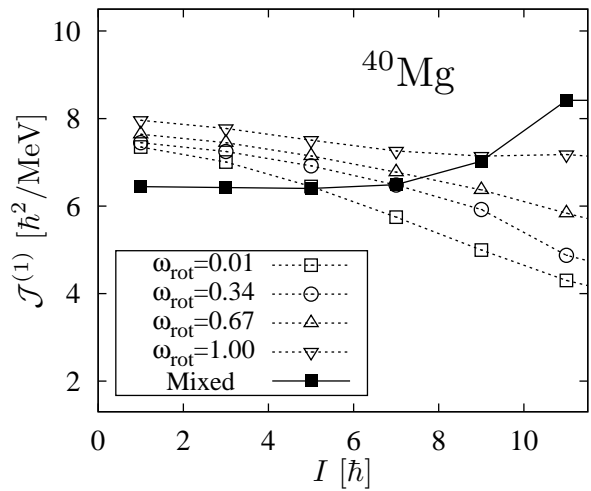}
\vspace*{-4mm}
\caption{
Moments of inertia versus spin value for $^{40}$Mg obtained by
the simple projection from one intrinsic HFB state
with four values of the cranking frequency,
$\hbar\omega_{\rm rot}=0.01,0.34,0.67,1.00$~MeV,
compared with the result of the projected configuration mixing
employing those four HFB states.
}
\label{fig:MgMoIm4}
\end{center}
\end{figure}

In Figs.~\ref{fig:MgMoIm} and~\ref{fig:MgMoIm4},
we compare the results of the moment of inertia ${\cal J}^{(1)}$
obtained by various calculations as functions of spin.
The inertia by the simple projection from the non-cranked HFB state
is smaller than the others and clearly shows the importance of the effect
of cranking to increase the moment of inertia.  However, the effect is
not so conspicuous as in the case of $^{164}$Er;
the calculated moment of inertia by the simple projection from
the non-cranked HFB state is about 80\% of that by
the projected configuration mixing.
It is interesting to mention that all inertias calculated by
the simple projection from one HFB state decrease as functions of spin
in this nucleus; see Fig.~\ref{fig:MgMoIm4}.
Especially the reduction of the result with the smallest cranking frequency,
$\hbar\omega_{\rm rot}=0.01$~MeV, is largest.
This kind of behavior is quite unusual and
it was speculated in Ref.~\cite{TSFSD14} that this reduction of inertia
may be characteristic for the unstable skin nucleus.
As it is shown in Figs.~\ref{fig:MgMoIm} and~\ref{fig:MgMoIm4}, however,
the configuration mixing recovers the usual behavior of the moment of inertia:
It is almost constant or gradually increases at higher spins.
We believe that the result of the configuration mixing is more reliable
and the decrease of the moment of inertia
as a function of spin~\cite{TSFSD14} would not be realized.
Although the reduction of inertia seems to be unrealistic,
it is still interesting to investigate the rotational motion
in the deformed unstable nuclei up to the neutron drip-line.
A systematic Gogny HFB and the projection calculations for the Mg isotopes
including $^{40}$Mg are now in progress~\cite{MgWat}.

\begin{figure}[!htb]
\begin{center}
\includegraphics[width=155mm]{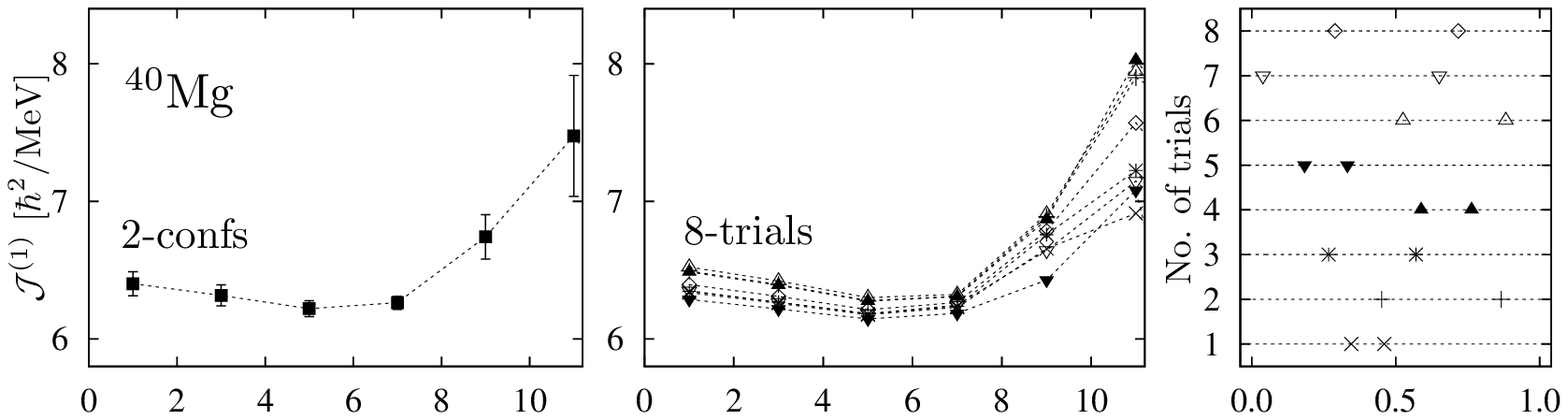}
\includegraphics[width=155mm]{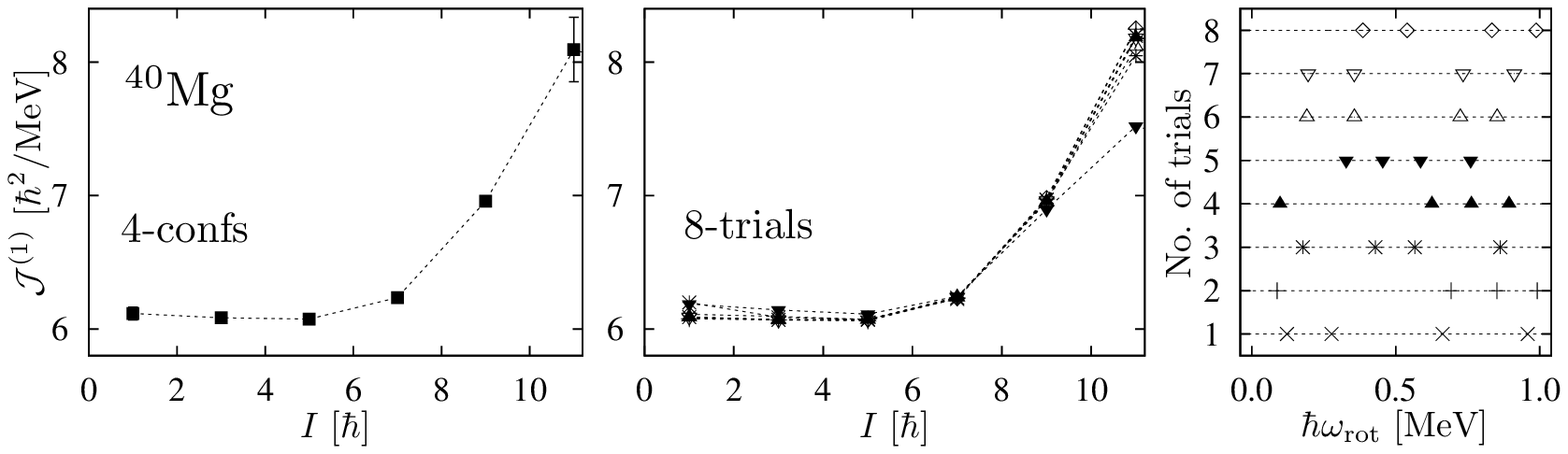}
\vspace*{-4mm}
\caption{
Analysis of moments of inertia for $^{40}$Mg calculated by
the projected configuration mixing employing cranked HFB states
with randomly chosen cranking frequencies.
The smaller model space $N_{\rm osc}^{\rm max}=8$ is used in this calculation.
Two sets of three panels in the first and second raw show
the results of calculation with mixing two and four configurations,
respectively.
In each set the left panel depicts the average moment of inertia
with the error bars indicating the standard deviation for eight trials of
randomly chosen frequencies, which are shown in the right panel as crosses,
and the middle panel displays the moments of inertia calculated
in those eight trials.
The results of eight trials are distinguished by different symbols.
}
\label{fig:MgRmoi}
\end{center}
\end{figure}

In order to see how the result of configuration mixing depends on the choice
of cranking frequencies, similar analysis to the case of $^{164}$Er
has been performed for $^{40}$Mg by using randomly chosen frequencies;
the result is shown in Fig.~\ref{fig:MgRmoi}.
In this calculation the smaller basis size with $N_{\rm osc}^{\rm max}=8$
is used to reduce the numerical task because the configuration mixing
calculations should be repeated many times.
The random numbers in this case are generated in the range,
$0<\hbar\omega_{\rm rot}<1.0$~MeV, and with the condition that
the difference of two nearest frequencies satisfies,
$\hbar\Delta\omega_{\rm rot}>0.1$~MeV,
to avoid the large overlap between the two associated HFB states.
The results with the two and four configurations are shown
in the first and second raw, respectively, in Fig.~\ref{fig:MgRmoi}.
The eight sets of cranking frequencies are generated for trials,
which are shown in the right panels of the figure,
and the eight moments of inertia calculated by
the projected configuration mixing in each trial are displayed
in the middle panels.   The average of these eight trials is shown
in the left panels with the error bars being the standard deviation;
in fact the standard deviations are rather small except at the highest spin.
As it is clear, the mixing with four configurations gives
rather similar results for all the eight trials,
which is in contrast to the case of $^{164}$Er,
where the standard deviations are small only
in the result with mixing five configurations.
The difference between the converged result of mixing
the four configurations in Fig.~\ref{fig:MgRmoi}
and that in Fig.~\ref{fig:MgMoIm} is due to
the different model space with $N_{\rm osc}^{\rm max}=8$ and~12 being used.
The large standard deviations at the highest spin in this example may be
related to the irregularity corresponding to the gradual alignments
at higher rotational frequency in Fig.~\ref{fig:MgIom}.

%###################################################
\subsection{Superdeformed band of \,$^{\it 152}$\hspace*{-2.5pt}Dy}
\label{sec:Dy-SD}

\begin{figure}[!htb]
\begin{center}
\includegraphics[width=70mm]{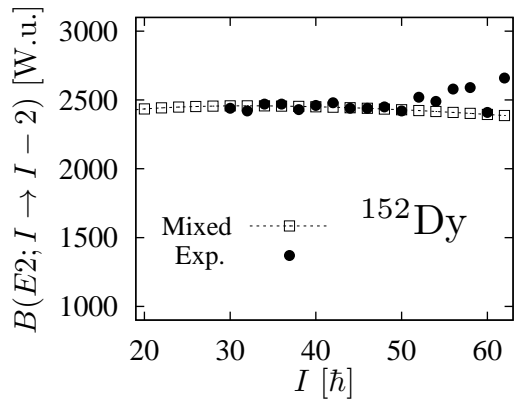}
\vspace*{-4mm}
\caption{
Calculated in-band $B(E2;I\rightarrow I-2)$ values
by the projected configuration mixing in for the superdeformed state band
of $^{152}$Dy in comparison with the experimental data.
}
\label{fig:DyBE2}
\end{center}
\end{figure}

The last example is the superdeformed band of a nucleus $^{152}$Dy.
It is the first nucleus, in which the high-spin superdeformed rotational band
was discovered by the discrete-line gamma-ray spectroscopy~\cite{Twin86}.
Moreover, the linking transitions between the superdeformed
and normal deformed states were measured in this nucleus~\cite{Dy152SD},
and the excitation energy and the precise spin-assignment
of the superdeformed band are known.
We have done the HFB and the projection calculations using
the oscillator basis with $N_{\rm osc}^{\rm max}=12$ for this example.
It has been found that the superdeformed minimum exists
in the non-cranked HFB calculation, where the neutron pairing vanishes
but the very weak proton pairing remains
with the average pairing gap $\bar{\Delta}_\pi\approx 0.4$~MeV.
The projection calculation with the pairing correlation is much more
time consuming~\cite{TS12} and we need to calculate up to very high-spin
states like $I\approx 60$.
Therefore we switched off the proton pairing correlation
for the calculation of this example.
The difference between the minimum of the binding energies
with and without the proton pairing correlation is
less than 20~keV within the mean-field approximation.
The selfconsistent deformation is axially-symmetric
and the quadrupole deformation parameter is $\beta_2=0.715$,
which decreases gradually up to 0.699
when the system is cranked up to $\hbar\omega_{\rm rot}=0.7$~MeV
keeping very well the axial-symmetry.
This result of the mean-field calculation well corresponds
to those of the Woods-Saxon- or Nilsson-Strutinsky calculations
of, e.g., Refs.~\cite{NWJ89,SVB90}, although the magnitude of deformation
is slightly larger than the results of these references.
This difference is due to the different definition of the deformation;
in the Strutinsky calculation the deformation is defined for the average
potential, while in the HFB calculation it is defined by the density
distribution, see Ref.~\cite{NWJ89}.
In fact the calculated deformation well reproduces the measured quadrupole
moment, see the end of this subsection related to Fig.~\ref{fig:DyBE2}.
As for the projection calculation we take $I_{\rm max}=62$ and $K_{\rm max}=22$
and therefore the numbers of integration mesh points are
$N_\alpha=N_\gamma=46$ and $N_\beta=126$ for the Euler angles.

In this example we do not show either the rotational spectrum nor
the angular momentum versus rotational frequency relation,
but rather we concentrate on the moment of inertia.
The projected configuration mixing calculation has been performed with
four cranked HFB states with the equidistant cranking frequencies,
$(\hbar\omega_{\rm rot}^{(n)},\,n=1:4)$
=$(0.01,\,0.24,\,0.47,\,0.70)$~MeV.
with the value of norm cut-off taken to be $10^{-9}$.
The energy gain of the $0^+$ superdeformed state by the simple projection
from the non-cranked HFB state is 4.11~MeV
and that by the projected configuration mixing is 4.14~MeV;
the difference is rather small in this case.
The calculated excitation energy of the superdeformed $24^+$ state
by the configuration mixing is $9.97$~MeV,
which is compared to the experimentally measured value
$10.644$~MeV~\cite{Dy152SD}; the agreement is rather nice.

\begin{figure}[!htb]
\begin{center}
\includegraphics[width=70mm]{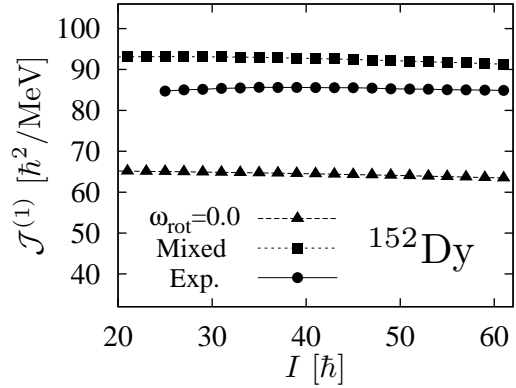}
\vspace*{-4mm}
\caption{
Moments of inertia versus spin value for $^{152}$Dy obtained by
the two calculations, the projection from the non-cranked HFB state
(``$\omega_{\rm rot}=0.0$'') and
and the projected configuration mixing (``Mixed'')
in comparison with the experimental data (``Exp.'').
The oscillator basis with $N_{\rm max}=12$ is used.
}
\label{fig:DyMoIm}
\end{center}
\end{figure}

\begin{figure}[!htb]
\begin{center}
\includegraphics[width=70mm]{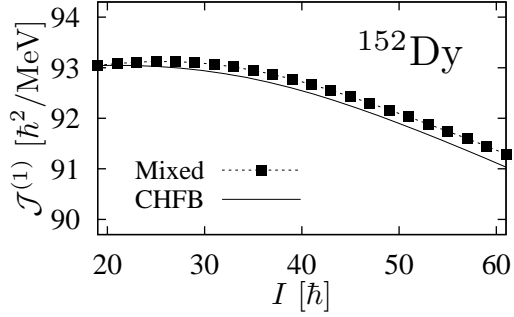}
\vspace*{-4mm}
\caption{
Moments of inertia versus spin value for $^{152}$Dy obtained by
the cranked HFB (the label ``CHFB'') is compared with
the result of the projected configuration mixing (``Mixed'').
The ordinate scale is enlarged compared with Fig.~\ref{fig:DyMoIm}
to see the difference.
}
\label{fig:DyMoIc}
\end{center}
\end{figure}

The moments of inertia for various calculations and experimental data
are displayed in Figs.~\ref{fig:DyMoIm} and~\ref{fig:DyMoIc}, separately.
As is clearly seen, all the calculated inertias are rather constant
as functions of spin in accordance with the experimental data,
although the absolute value of the result of the projected configuration mixing
is slightly larger than the experimental one.
On the other hand the result of the simple projection
from the non-cranked HFB state is considerably smaller than
the experimental one, and again indicates the importance of
the time-odd components in the wave function induced by the cranking term.
The semiclassical moment of inertia of the cranked HFB calculated
in the same way as in the previous examples coincides very well
with that of the projected configuration mixing in this example;
the ordinate scale is extremely enlarged to see the difference
in Fig.~\ref{fig:DyMoIc}.

\begin{figure}[!htb]
\begin{center}
\includegraphics[width=80mm]{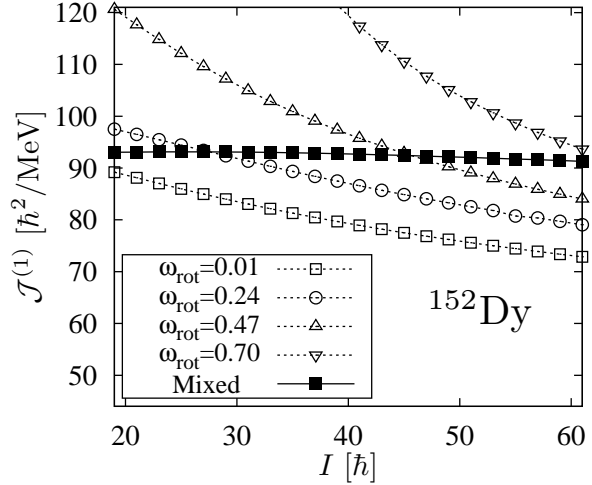}
\vspace*{-4mm}
\caption{
Moments of inertia versus spin value for $^{152}$Dy obtained by
the simple projection from one intrinsic HFB state
with four values of the cranking frequencies,
$\hbar\omega_{\rm rot}=0.01,\,0.24,\,0.47,\,0.70$~MeV, respectively,
compared with the result of the projected configuration mixing
employing those four HFB states.
}
\label{fig:DyMoIm4}
\end{center}
\end{figure}

\begin{figure}[!htb]
\begin{center}
\includegraphics[width=70mm]{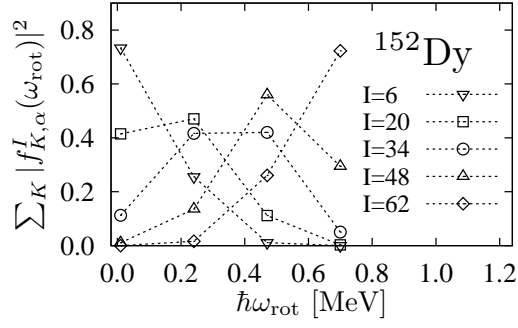}
\vspace*{-4mm}
\caption{
Probability distribution (Eq.~(\ref{eq:probf2}))
over the four HFB configurations with different cranking frequencies
for the spin $I$ member of the superdeformed rotational band of $^{152}$Dy
obtained by the projected configuration mixing.
}
\label{fig:Dyprob}
\end{center}
\end{figure}

In order to show the effect of configuration mixing,
we depict in Fig.~\ref{fig:DyMoIm4} four moments of inertia
obtained by the simple projection from one cranked HFB state
with different frequencies,
$\hbar\omega_{\rm rot}=0.01,\,0.24,\,0.47,\,0.70$~MeV,
in addition to the result of the projected configuration mixing.
Interestingly enough the calculated inertias by the simple projection
are considerably different,
both their absolute values and spin-dependences,
as is clearly seen from the figure;
the one associated with the higher cranking frequency
has larger absolute value and more strongly decreases as a function of spin.
This is in contrast to the previous two examples of
the ground state bands in $^{164}$Er and $^{40}$Mg.
Although the strong reduction of these inertias
completely contradicts with the experimental data,
the projected configuration mixing makes the resultant inertia
almost constant in accordance with the tendency of
the experimentally measured inertia.
This fact clearly shows again the importance of the configuration mixing
to obtain the correct rotational behavior of the superdeformed states
by the angular momentum projection method.
In order to study how the four configurations mix,
the probability distributions are depicted in Fig.~\ref{fig:Dyprob}.
In this case the distributions are just what are expected, i.e.,
the probabilities have nice peaks at the cranking frequencies corresponding to
$ \langle\Phi_{\rm cr}(\omega_{\rm rot})|
 J_y|\Phi_{\rm cr}(\omega_{\rm rot})\rangle \approx I\,\hbar$,
in contrast to the case of $^{164}$Er.
This clearly shows that the superdeformed rotational band
can be interpreted as a very good semiclassical rotor.

\begin{figure}[!htb]
\begin{center}
\includegraphics[width=70mm]{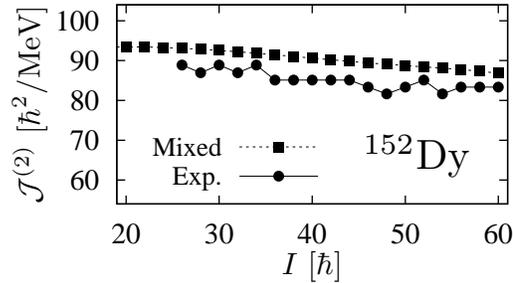}
\vspace*{-4mm}
\caption{
The second moment of inertia versus spin value for $^{152}$Dy obtained by
the projected configuration mixing (``Mixed'')
in comparison with the experimental data (``Exp.'');
see the text for the precise definition of ${\cal J}^{(2)}$.
}
\label{fig:DyMoI2m}
\end{center}
\end{figure}

Figure~\ref{fig:DyMoI2m} depicts the result of configuration mixing
for the so-called second (or dynamic) moment of inertia~\cite{BM81},
defined by ${\cal J}^{(2)}(I) \equiv 4\hbar^2/(E(I+2)+E(I-2)-2E(I))$,
in comparison with the experimental data.
The spin-assignment is not necessary to estimate this inertia
and therefore it has been used frequently
for the study of high-spin superdeformed bands.
As it is seen in the figure, a better agreement with the experimental data
is obtained compared with the first moment of inertia in Fig.~\ref{fig:DyMoIm},
although the calculated inertia is still slightly overestimated.

\begin{figure}[!htb]
\begin{center}
\includegraphics[width=155mm]{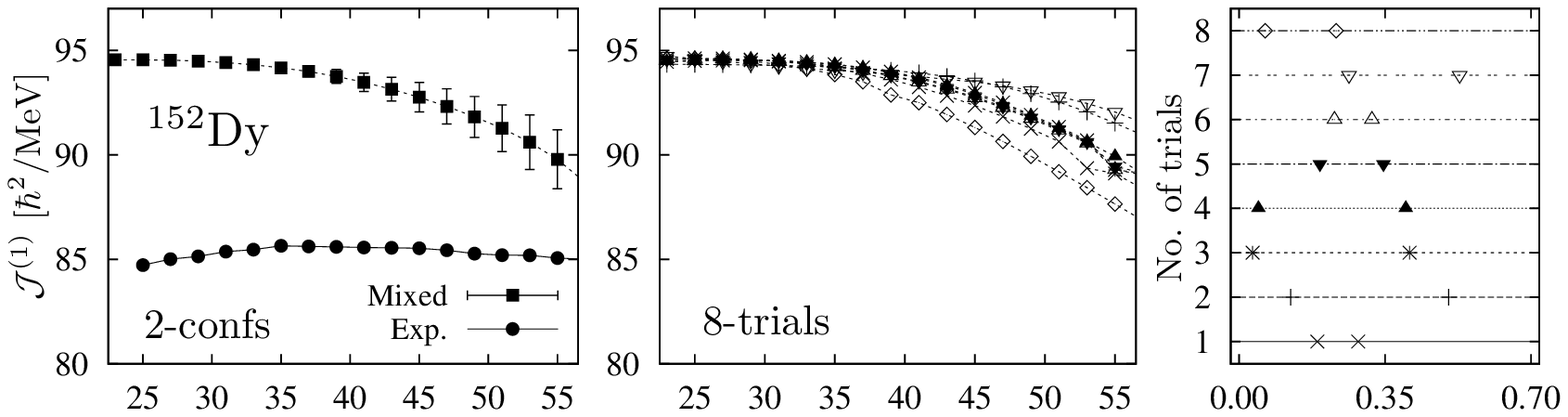}
\includegraphics[width=155mm]{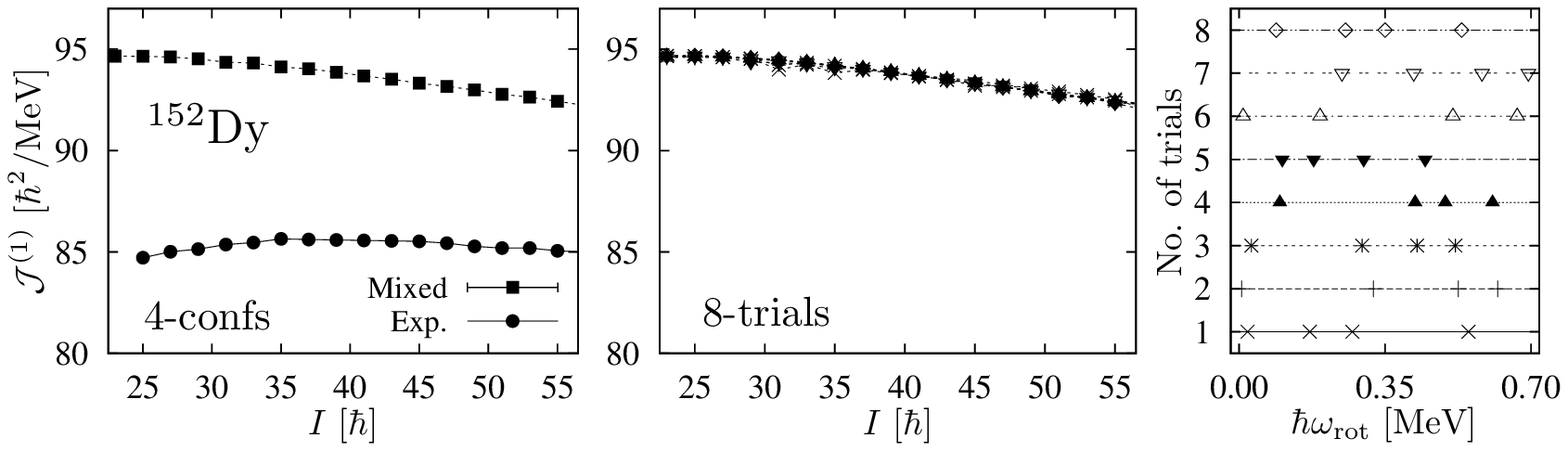}
\vspace*{-4mm}
\caption{
Analysis of moments of inertia for $^{152}$Dy calculated by
the projected configuration mixing employing cranked HFB states
with randomly chosen cranking frequencies.
The smaller model space $N_{\rm osc}^{\rm max}=10$ is used in this calculation.
Two sets of three panels in the first and second raw show
the results of calculation with mixing two and four configurations,
respectively, in comparison with the experimental data.
In each set the left panel depicts the average moment of inertia
with the error bars indicating the standard deviation for eight trials of
randomly chosen frequencies, which are shown in the right panel as crosses,
and the middle panel displays the moments of inertia calculated
in those eight trials.
The results of eight trials are distinguished by different symbols.
}
\label{fig:DyRmoi}
\end{center}
\end{figure}

It is interesting to see how the result of configuration mixing depends on
the choice of cranking frequencies also in this case.
Similar analysis to the cases of $^{164}$Er and $^{40}$Mg by employing
the randomly chosen sets of cranking frequencies has been performed and
the result is shown in Fig.~\ref{fig:DyRmoi}.
In this calculation the smaller basis size with $N_{\rm osc}^{\rm max}=10$
is used to reduce the numerical task.
The random numbers in this case are generated in the range,
$0<\hbar\omega_{\rm rot}<0.7$~MeV, and with the condition that
the difference of two nearest frequencies satisfies,
$\hbar\Delta\omega_{\rm rot}>0.07$~MeV.
The eight sets of cranking frequencies are employed,
and the results with the two and four configurations are shown
in the first and second raw, respectively, in Fig.~\ref{fig:DyRmoi}.
The left panels depict the average of eight trials
with the error bars being the standard deviation, the middle panels
the eight moments of inertia calculated in each trial,
and the right panels the generated sets of frequencies.
Surprisingly the mixing with only two configurations already gives
a rather converging result, and the use of four configurations
is enough to attain the almost completely unique result:
Note that the ordinate scale is very enlarged in Fig.~\ref{fig:DyRmoi}
compared with Fig.~\ref{fig:DyMoIm}.
The error bars are invisible in the left-bottom panel and the results
of eight trials cannot be distinguished even though the eight sets of
randomly chosen frequencies are considerably different.
This result clearly confirms again that a rather small number of configurations,
four in this case, is sufficient to obtain reliable results.
The difference of about 2\% between the result of mixing
the four configurations in Fig.~\ref{fig:DyRmoi}
and that in Fig.~\ref{fig:DyMoIm} is due to
the different model space with $N_{\rm osc}^{\rm max}=10$ and~12 being used.

\begin{figure}[!htb]
\begin{center}
\includegraphics[width=70mm]{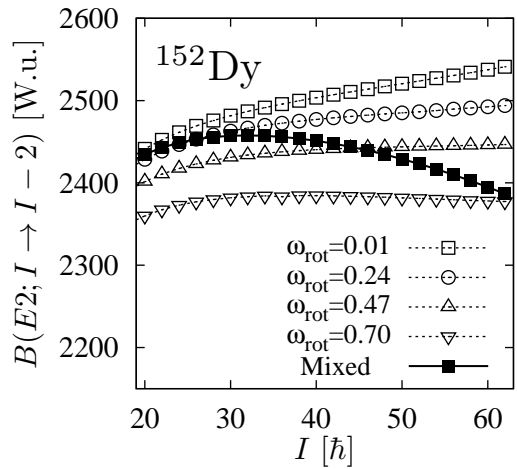}
\vspace*{-4mm}
\caption{
Calculated in-band $B(E2;I\rightarrow I-2)$ values
by the simple projection from one HFB state with different cranking
frequencies in comparison with the result of configuration mixing.
The ordinate scale is enlarged compared with Fig.~\ref{fig:DyBE2}.
}
\label{fig:DyBE2m5}
\end{center}
\end{figure}

Finally we compared the calculated $B(E2)$ values with
the experimental data~\cite{Dy152BE2} in Fig.~\ref{fig:DyBE2}.
Although the increasing trend of the experimental data at higher spins,
$I > 50$, is not reproduced, nice agreement is obtained.
This indicates that the selfconsistently calculated quadrupole deformation
agrees very well with the experimental data
at least at the spin region, $I \le 50$.
In order to show the effect of configuration mixing for the $B(E2)$ values,
those calculated by the simple projection from one HFB state
with different cranking frequencies are depicted in Fig.~\ref{fig:DyBE2m5}.
As it clear from Fig.~\ref{fig:DyBE2m5}, the slight decrease of
the calculated $B(E2)$ in the spin range, $I > 30$,
is a result of the configuration mixing.
This can be understood as follows: The selfconsistently calculated
deformation parameter $\beta_2$ decreases as a function of
the cranking frequency with keeping very well the axial-symmetry
as was already mentioned, which leads to the result that
the $B(E2)$ value calculated by the simple projection is smaller
when projected from the HFB state with higher frequency
as is also shown in Fig.~\ref{fig:DyBE2m5}.
The mixing probabilities in this case have sharp peaks at the frequencies
corresponding to the spin values under consideration, see Fig.~\ref{fig:Dyprob}.
Therefore the quadrupole moment calculated by the configuration mixing
effectively depends on the spin and decreases as a function of it.
Namely, the result of the configuration mixing directly reflects
the selfconsistent change of the mean-field deformation as increasing
the cranking frequency, which is in contrast to the case of $^{164}$Er,
where the mixing probabilities spread and the calculated $B(E2)$ values
by the configuration mixing do not directly reflect
the selfconsistent change of the mean-field deformation.

%------------------------------------------------------------------------------
\section{Summary and Discussion}
\label{sec:summary}

In the present work we propose a method to reliably calculate
the rotational band by employing the fully microscopic
angular momentum projection technique combined with the cranking model.
The method is based on the configuration mixing after the projection,
in the sense of the GCM, employing several cranked HFB states
with different rotational frequencies;
the original idea was suggested by Peierls-Thouless~\cite{PT62}.
By applying the method to a few realistic examples,
it has been found that the necessary number of cranked HFB states employed
in the configuration mixing is rather small, something like four to five
in the examples studied.   The result is essentially independent of
the chosen set of cranking frequencies and the set of equidistant frequencies
in a suitably chosen range has been mainly utilized.

One of the examples of application is
the ground state rotational band of a typical rare-earth nucleus $^{164}$Er;
with the Gogny D1S interaction,
a reasonably good agreement with the experimental data
has been obtained for both the spectrum and $B(E2)$
in the whole spin range, $0 \le I \le 22$.
Especially, the increase of the moment of inertia as a function of spin
is almost reproduced as a result of the configuration mixing,
while the calculated inertias by the simple projection
from one cranked HFB state with various frequencies
are almost constant.
Note that there are no adjustable parameters in this calculation.
Another example is the superdeformed rotational band in $^{152}$Dy;
the fairly constant moment of inertia in a wide spin range
appears as a result of the configuration mixing
in accordance with the experimental data, although the agreement
is not perfect because the selfconsistent deformation is slightly overestimated.
The calculated inertias by the simple projection from one cranked HFB state
with various frequencies decrease as spin increases in this case;
the one with the non-cranked HFB state is almost constant
but its absolute value is only about 70\% of the calculated inertia
by the projected configuration mixing.
The feature that the calculated inertias decrease is also observed
in the ground state rotational band of the unstable neutron-skin nucleus
$^{40}$Mg if calculated by the simple projection from one cranked HFB state.
In these two cases, the superdeformed band of $^{152}$Dy and
the ground state band of $^{40}$Mg, the pairing correlation vanishes
for both neutrons and protons;
the reduction of inertia may be related to this feature.

In all three examples, the semiclassical cranking moment of inertia
is compared with the calculated inertia by the projected configuration mixing.
In the superdeformed band of $^{152}$Dy and the ground state band of $^{40}$Mg
the two inertias agrees quite well.  In the ground state band of $^{164}$Er,
however, the cranking inertia more rapidly increases than
the one calculated by the projected configuration mixing,
and their difference reaches about 20\% at higher spins.
Thus only the results of the projected multi-cranked configuration mixing
are in accordance with the experimental data and
clearly indicate the importance of configuration mixing.

In Ref.~\cite{PT62} it was discussed that the superposition with respect to
both the position and velocity as generating coordinates is necessary
in order to reproduce the correct inertial mass for the center-of-mass motion;
the so-called ``double projection''~\cite{RS80}.
The prescription of Eq.~(\ref{eq:PTanz}) is its natural extension
to the collective rotational motion~\cite{PT62}.
Although the inertial mass for rotation is not known {\it a priori},
the result of the present work shows
that the inclusion of the effect of cranking,
i.e., superposing the wave functions with different angular velocities,
is very important.
In fact, as it was already stressed in Ref.~\cite{TS12},
the moment of inertia calculated by the simple projection
from the non-cranked (time-even) HFB state is too small.
The calculated inertia considerably increases if the projection is
performed from a cranked HFB state with small cranking frequency;
it is the crudest approximation to Eq.~(\ref{eq:PTanz}), i.e.,
the frequency integration is approximated by only one point.
When the number of points is increased the result of
the projected configuration mixing converges independently of
the chosen set of cranking frequencies,
which is natural because the procedure is a discrete approximation
of the frequency integration in Eq.~(\ref{eq:PTanz}).
What is surprising, however, is that
the necessary number of cranked HFB states is rather small
and the procedure is tractable,
which is an important consequence of the present work.

It may be worth mentioning that the behavior of convergence
when increasing the number of randomly chosen configurations for $^{164}$Er
is considerably different from those for $^{40}$Mg and $^{152}$Dy.
Moreover, the necessary number of configurations for convergence
is largest in the case of $^{164}$Er.
Although we do not understand the definite reason for this behavior,
it may be most probably related to the presence of pairing correlations.

We have shown that the superposition with respect to the cranking frequency
is a key for a reliable description of the collective rotation
by employing the angular momentum projection method.
The Projected Shell Model~\cite{HS95} is also based on the projection
and the configuration mixing, but does not rely on the cranking model:
It utilizes much larger but simple configurations in a sense of the shell model,
i.e., two, four, and, if necessary, higher numbers of quasiparticle excited
configurations associated with the deformed and superconducting mean-field.
We believe that the cranking model is an efficient way to incorporate
the multi-quasiparticle excitations suitable for
describing the collective rotation.
To understand the mutual relationship between
the Projected Shell Model and the multi-cranked configuration mixing
in the present work may be an interesting future problem.

%-------------------------------------------------------------------------------
\section*{Acknowledgements}

This work is supported in part
by Grant-in-Aid for Scientific Research (C) 
No.~25$\cdot$949 from Japan Society for the Promotion of Science.

%\newpage
\vspace*{10mm}

%-------------------------------------------------------------------------------

\end{document}